\documentstyle[aps]{revtex}

\begin{document}
\title{New closed expression of the interaction kernel in the Bethe-Salpeter
equation for quark-antiquark bound states}
\author{Jun-Chen Su}
\address{Center for Theoretical Physics, Physics College, Jilin University,\\
Changchun 130023, People's Republic of China}
\maketitle

\begin{abstract}
The interaction kernel in the Bethe-Salpeter equation for quark-antiquark
bound states is derived newly from QCD in the case where the quark and the
antiquark are of different flavors. The technique of the derivation is the
usage of the irreducible decomposition of the Green's functions involved in
the Bethe-Salpeter equation satisfied by the quark-antiquark four-point
Green's function. The interaction kernel derived is given a closed and
explicit expression which shows a specific structure of the kernel since the
kernel is represented in terms of the quark, antiquark and gluon propagators
and some kinds of quark, antiquark and/or gluon three, four, five and
six-point vertices. Therefore, the expression of the kernel is not only
convenient for perturbative calculations, but also suitable for
nonperturbative investigations.

PACS: 11.10St, 12.38.Aw, 12.38.Gc, 12.40.Pp

Key words: Bethe-Salpeter equation, interaction kernel, quark-antiquark
bound state.
\end{abstract}

\section{Introduction}

The Bethe-Salpeter (B-S) equation which was proposed early in Refs. [1, 2],
commonly, is recognized as a rigorous approach to the relativistic bound
state problem [1-13 ]. The distinctive features of the equation are: (1) The
equation is derived from the quantum field theory and hence set up on the
firm dynamical basis; (2) The interaction kernel in the equation contains
all the interactions taking place in the bound states and therefore the
equation provides a possibility of exactly solving the problem of
relativistic bound states; (3) The equation is elegantly formulated in a
manifestly Lorentz-covariant form in the Minkowski space which allows us to
discuss the equation in any coordinate frame. However, there are tremendous
difficulties in practical applications of the equation. One of the
difficulties arises from the fact that the kernel in the equation was not
given a closed form in the past. The kernel usually is defined as a sum of
B-S (two-particle) irreducible Feynman diagrams each of which can only be
individually determined by a perturbative calculation. This definition is,
certainly, not suitable to investigate the subjects such as the nuclear
force in nuclear physics and the quark confinement in hadron physics which
must necessarily be solved by a nonperturbative method. This is why as said
in Ref. [12] that'' The Bethe-Salpeter equation has not led to a real
breakthrough in our understanding of the quark-quark force''.

Opposite to the conventional concept as commented in Ref. [13] that ''The
kernel $K$ can not be given in closed form expression'', we have derived a
closed and compact expression of the B-S kernel for quark-antiquark ($q%
\overline{q}$) bound states in a recent publication [14]. The expression is
derived with the aid of equations of motion satisfied by the $q\overline{q}$
four-point Green's function and some other kinds of Green's functions and
contains only a few types of Green's functions which not only are easily
calculated by the perturbation method, but also suitable to be investigated
by a certain nonperturbation approach. In order to exhibit a more specific
structure of the B-S kernel, in this paper, we are devoted to deriving a new
expression of the kernel by means of the technique of irreducible
decomposition of the Green's functions involved in the B-S equation
satisfied by the $q\overline{q}$ four-point Green's function. The technique
was successfully applied to derive the interaction kernel in the
Dirac-Schr\"odinger equation for $q\overline{q}$ bound states in the
author's latest publication [15]. The new expression derived is represented
in terms of the quark, antiquark and gluon propagators as well as some kinds
of quark, antiquark and/or gluon three, four, five and six-line vertices.
For brevity of the derivation, we restrict ourself in this paper to discuss
the B-S kernel for the bound system consisting of a quark and an antiquark
which are of different flavors.

The rest of this paper is arranged as follows. In Sect. 2, we sketch the B-S
equation satisfied by the $q\overline{q}$ four-point Green's function. In
Sect. 3, we describe the B-S reducibility of the Green's functions involved
in the B-S equation by means of the technique of irreducible decomposition
of Green's functions. Section 4 is used to derive the final expression of
the B-S kernel given by the irreducible decomposition of the Green's
functions. The last section serves to make summary and some remarks. In
Appendix, we show some details of the irreducible decomposition which
follows from the QCD generating functional.

\section{B-S equation and definition of the interaction kernel}

In Ref. [14], it was shown that to derive the B-S interaction kernel for $q%
\overline{q}$ bound states, it is necessary to derive a B-S equation
satisfied by the $q\overline{q}$ four-point Green's function. In the case of
quark and antiquark with different flavors, the $q\overline{q}$ four-point
Green's function is, in the Heisenberg picture, defined as [16]

\begin{equation}
G(x_{1,}x_2;y_1,y_2)_{\alpha \beta \rho \sigma }=\langle 0^{+}\mid T\{{\bf %
\psi }_\alpha (x_1){\bf \psi }_\beta ^c(x_2)\overline{{\bf \psi }}_\rho (y_1)%
\overline{{\bf \psi }}_\sigma ^c(y_2)\}\mid 0^{-}\rangle  \eqnum{1}
\end{equation}
where $T$ stands for the time-ordering product, ${\bf \psi }_\alpha (x)$ and 
${\bf \psi }_\beta ^c(x)$ are the quark and antiquark field operators,
respectively, $\overline{{\bf \psi }}_\alpha (x)$ and $\overline{\text{ }%
{\bf \psi }}_\beta ^c(x)$ are their corresponding Dirac conjugates, 
\begin{equation}
{\bf \psi }^c(x)=C\overline{{\bf \psi }}^T(x),\overline{{\bf \psi }^c}(x)=-%
{\bf \psi }^T(x)C^{-1}  \eqnum{2}
\end{equation}
here $C$ is the charge conjugation operator. The advantage of using $\psi
^c(x)$ other than $\overline{\psi }(x)$ to represent the antiquark field is
that the antiquark would behave as a quark in the B-S equation so that the
quark-antiquark equation formally is the same as the corresponding two-quark
equation.

For deriving the B-S equation satisfied by the $q\overline{q}$ four-point
Green's function , it is necessary to use various equations of motion
satisfied by the quark and antiquark propagators, the $q\overline{q}$
four-point Green's functions and some other kinds of Green's functions. The
latter equations of motion are easily derived from the QCD generating
functional [16] 
\begin{equation}
Z[J,\overline{\eta },\eta ,\overline{\xi },\xi ]=\frac 1N\int {\cal D(}A,%
\overline{\psi },\psi ,\overline{C},C)e^{iI}  \eqnum{3}
\end{equation}
where 
\begin{equation}
I=\int d^4x[{\cal L}+J^{a\mu }A_\mu ^a+\overline{\eta }\psi +\overline{\psi }%
\eta +\overline{\xi }C+\overline{C}\xi ]  \eqnum{4}
\end{equation}
in which ${\cal L}$ is the effective Lagrangian of QCD 
\begin{equation}
{\cal L}=\overline{\psi }(i{\bf \partial -}m+g{\bf A)}\psi -\frac 14F^{a\mu
\nu }F_{\mu \nu }^a-\frac 1{2\alpha }(\partial ^\mu A_\mu ^a)^2+\overline{C}%
^a\partial ^\mu (D_\mu ^{ab}C^b)  \eqnum{5}
\end{equation}
here ${\bf \partial }_x{\bf =}\gamma ^\mu \partial _\mu ^x$, ${\bf A=}\gamma
^\mu T^aA_\mu ^a$ with A$_\mu ^a$ being the vector potentials of gluon
fields and $T^a=\lambda ^a/2$ the quark color matrix, 
\begin{equation}
F_{\mu \nu }^a=\partial _\mu A_\nu ^a-\partial _\nu A_\mu ^a+gf^{abc}A_\mu
^bA_\nu ^c  \eqnum{6}
\end{equation}
are the strength tensors of the gluon field, 
\begin{equation}
D_\mu ^{ab}=\delta ^{ab}\partial _\mu +gf^{abc}A_\mu ^bC^c  \eqnum{7}
\end{equation}
are the covariant derivatives, $\overline{C}^a,C^b$ are the ghost fields,
and $J^{a\mu },\overline{\eta },\eta ,\overline{\xi }$ and $\xi $ denote the
external sources coupled to the gluon, quark and ghost fields respectively.
By the charge conjugation transformations shown in Eq. (2) for the quark
fields and in the following for the external sources 
\begin{equation}
\eta ^c=C\overline{\eta }^T,\overline{\eta ^c}=-\eta ^TC^{-1}  \eqnum{8}
\end{equation}
it is easy to prove the following relation 
\begin{equation}
\overline{\psi }(i{\bf \partial -}m+g{\bf A)}\psi +\overline{\eta }\psi +%
\overline{\psi }\eta =\overline{\psi ^c}(i{\bf \partial -}m+g\overline{{\bf A%
}}{\bf )}\psi ^c+\overline{\eta ^c}\psi ^c+\overline{\psi ^c}\eta ^c 
\eqnum{9}
\end{equation}
where $\overline{{\bf A}}=\gamma ^\mu \overline{T}^aA_\mu ^a$ with $%
\overline{T}^a=-\lambda ^{a*}/2$ being the antiquark color matrix.

Now let us sketch the derivation of the B-S equation satisfied by the $q%
\overline{q}$ four-point Green's function. The equation may be set up by
acting on the Green's function with the operator $(i{\bf \partial }_{x_1}%
{\bf -}m_1{\bf +}$ $\Sigma )(i{\bf \partial }_{x_2}{\bf -}m_2{\bf +}$ $%
\Sigma ^c)$ where, $m_{1\text{ }}$ and $m_2$ are the quark and antiquark
masses, and $\Sigma $ and $\Sigma ^c$ stand for the quark and antiquark
proper self-energies. For this purpose, we first need to derive the
equations of motion for the four-point Green's function. As shown in Ref.
[14], when we take the successive functional derivatives of the generating
functional in Eq. (3) with respect to the field function $\overline{\psi }%
_\alpha (x_1)$ and sources $\overline{\eta }_\beta ^c(x_2),\eta _\rho (y_1)$
and $\eta _\sigma ^c(y_2)$ and finally setting all the sources to vanish, it
can be found that 
\begin{equation}
\begin{tabular}{l}
$\lbrack (i{\bf \partial }_{x_1}-m_1+\Sigma )G]_{\alpha \beta \rho \sigma
}(x_{1,}x_2;y_1,y_2)=\delta _{\alpha \rho }\delta
^4(x_1-y_1)S_F^c(x_2-y_2)_{\beta \sigma }$ \\ 
$-(\Gamma ^{a\mu })_{\alpha \gamma }G_\mu ^a(x_1\mid
x_1,x_2;y_1,y_2)_{\gamma \beta \rho \sigma }+\int d^4z_1\Sigma
(x_1,z_1)_{\alpha \gamma }G(z_1,x_2;y_1,y_2)_{\gamma \beta \rho \sigma }.$%
\end{tabular}
\eqnum{10}
\end{equation}
Similarly, successively differentiating the generating functional in Eq. (3)
with respect to the field $\overline{\psi }_\beta ^c(x_2)$ and sources $%
\overline{\eta }_\alpha (x_1),\eta _\rho (y_1)$ and $\eta _\sigma ^c(y_2)$,
and finally turning off all the sources, one may obtain 
\begin{equation}
\begin{tabular}{l}
$\lbrack (i{\bf \partial }_{x_2}-m_2+\Sigma ^c)G]_{\alpha \beta \rho \sigma
}(x_{1,}x_2;y_1,y_2)=\delta _{\beta \sigma }\delta
^4(x_2-y_2)S_F(x_1-y_1)_{\alpha \rho }$ \\ 
$-(\overline{\Gamma }^{b\nu })_{\beta \lambda }G_\nu ^b(x_2\mid
x_1,x_2;y_1,y_2)_{\alpha \lambda \rho \sigma }+\int d^4z_2\Sigma
^c(x_2,z_2)_{\beta \lambda }G(x_1,z_2;y_1,y_2)_{\alpha \lambda \rho \sigma
}. $%
\end{tabular}
\eqnum{11}
\end{equation}
In Eqs. (10) and (11), $m_1$ and $m_2$ are the quark and antiquark masses, $%
S_F(x_1-y_1)$ and $S_F^c(x_2-y_2)$ are the quark and antiquark propagators, $%
\Sigma (x,y)$ and $\Sigma ^c(x.y)$ represent the quark and antiquark
self-energies, respectively, 
\begin{equation}
(\Omega ^{a\mu })_{\alpha \gamma }=g(\gamma ^\mu T^a)_{\alpha \gamma },\text{
}(\overline{\Omega }^{b\nu })_{\beta \lambda }=g(\gamma ^\nu \overline{T}%
^b)_{\beta \lambda }  \eqnum{12}
\end{equation}
and 
\begin{equation}
\begin{tabular}{l}
$G_{\mu .}^a(x_i\mid x_1,x_2;y_1,y_2)_{\alpha \beta \rho \sigma }$ \\ 
$=\langle 0^{+}\mid T[{\bf A}_\mu ^a(x_i){\bf \psi }_\alpha (x_1){\bf \psi }%
_\beta ^c(x_2)\overline{{\bf \psi }}_\rho (y_1)\overline{{\bf \psi }}_\sigma
^c(y_2)]\mid 0^{-}\rangle $%
\end{tabular}
\eqnum{ 13}
\end{equation}
here $i=1,2$. Acting on Eq. (11) with the operator ( $i{\bf \partial }_{x_1}%
{\bf -}m_1{\bf +}$ $\Sigma $ ) ( or equivalently, acting on Eq. (10) with ($i%
{\bf \partial }_{x_2}{\bf -}m_2{\bf +}$ $\Sigma ^c$)) and utilizing the
equations satisfied by the quark and antiquark propagators 
\begin{equation}
\begin{array}{c}
\lbrack (i{\bf \partial }_{x_1}-m_1+\Sigma )S_F]_{\alpha \rho
}(x_1,y_1)=\delta _{\alpha \rho }\delta ^4(x_1-y_1), \\ 
\lbrack (i{\bf \partial }_{x_2}-m_2+\Sigma ^c)S_F^c]_{\beta \sigma
}(x_2,y_2)=\delta _{\beta \sigma }\delta ^4(x_2-y_2)
\end{array}
\eqnum{14}
\end{equation}
where 
\begin{equation}
\begin{array}{c}
(\Sigma S_F)_{\alpha \rho }(x_1,y_1)\equiv \int d^4z_1\Sigma
(x_1,z_1)_{\alpha \gamma }S_F(z_1-y_1)_{\gamma \rho }=(\Omega ^{a\mu
})_{\alpha \gamma }\Lambda _\mu ^a(x_1\mid x_1,y_1)_{\gamma \rho }, \\ 
(\Sigma ^cS_F^c)_{\beta \sigma }(x_2,y_2)\equiv \int d^4z_2\Sigma
^c(x_2,z_2)_{\beta \lambda }S_F^c(z_2-y_2)_{\lambda \sigma }=(\overline{%
\Omega }^{b\nu })_{\beta \lambda }\Lambda _\nu ^{{\bf c}b}(x_2\mid
x_2,y_2)_{\lambda \sigma }
\end{array}
\eqnum{15}
\end{equation}
in which 
\begin{equation}
\begin{array}{c}
\Lambda _\mu ^a(x_1\mid x_1,y_1)_{\gamma \rho }=\frac 1i\langle 0^{+}\mid T[%
{\bf A}_\mu ^a(x_1){\bf \psi }_\gamma (x_1)\overline{{\bf \psi }}_\rho
(y_1)]\mid 0^{-}\rangle , \\ 
\Lambda _\nu ^{{\bf c}b}(x_2\mid x_2,y_2)_{\lambda \sigma }=\frac 1i\langle
0^{+}\mid T[{\bf A}_\nu ^b(x_2){\bf \psi }_\lambda ^c(x_2)\overline{{\bf %
\psi }}_\sigma ^c(y_2)]\mid 0^{-}\rangle
\end{array}
\eqnum{16}
\end{equation}
and the equation obeyed by Green's function $G_\nu ^b(x_2\mid
x_{1,}x_2;y_1,y_2)$%
\begin{equation}
\begin{tabular}{l}
$\lbrack (i{\bf \partial }_{x_1}-m_1+\Sigma )G_\nu ^b]_{\alpha \lambda \rho
\sigma }(x_2\mid x_{1,}x_2;y_1,y_2)=\delta _{\alpha \rho }\delta
^4(x_1-y_1)\Lambda _\nu ^{{\bf c}b}(x_2\mid x_2,y_2)_{\lambda \sigma }$ \\ 
$-(\Gamma ^{a\mu })_{\alpha \gamma }G_{\mu \nu }^{ab}(x_1,x_2\mid
x_1,x_2;y_1,y_2)_{\gamma \lambda \rho \sigma }+\int d^4z_1\Sigma
(x_1,z_1)_{\alpha \gamma }G_\nu ^b(x_2\mid z_1,x_2;y_1,y_2)_{\gamma \lambda
\rho \sigma },$%
\end{tabular}
\eqnum{17}
\end{equation}
one may derive 
\begin{equation}
\begin{tabular}{l}
$\lbrack (i{\bf \partial }_{x_1}-m_1+\Sigma )(i{\bf \partial }%
_{x_2}-m_2+\Sigma ^c)G]_{\alpha \beta \rho \sigma }(x_{1,}x_2;y_1,y_2)$ \\ 
$=\delta _{\alpha \rho }\delta _{\beta \sigma }\delta ^4(x_1-y_1)\delta
^4(x_2-y_2)$ \\ 
$+(\Omega ^{a\mu })_{\alpha \gamma }(\overline{\Omega }^{b\nu })_{\beta
\lambda }G_{\mu \nu }^{ab}(x_1,x_2\mid x_1,x_2;y_1,y_2)_{\gamma \lambda \rho
\sigma }$ \\ 
$-(\Omega ^{a\mu })_{\alpha \gamma }\int d^4u_2\Sigma ^c(x_2,u_2)_{\beta
\lambda }G_\mu ^a(x_1\mid x_1,u_2;y_1,y_2)_{\gamma \lambda \rho \sigma }$ \\ 
$-(\overline{\Omega }^{b\nu })_{\beta \lambda }\int d^4u_1\Sigma
(x_1,u_1)_{\alpha \gamma }G_\nu ^b(x_2\mid u_1,x_2;y_1,y_2)_{\gamma \lambda
\rho \sigma }$ \\ 
$\int d^4z_1d^4z_2\Sigma (x_1,z_1)_{\alpha \gamma }\Sigma ^c(x_2,z_2)_{\beta
\lambda }G(z_1,z_2;y_1,y_2)_{\gamma \lambda \rho \sigma }$%
\end{tabular}
\eqnum{18}
\end{equation}
where 
\begin{equation}
\begin{tabular}{l}
$G_{\mu \nu }^{ab}(x_i,y_j\mid x_1,x_2;y_1,y_2)_{\alpha \beta \rho \sigma }$
\\ 
$=\langle 0^{+}\mid T[{\bf A}_\mu ^a(x_i){\bf A}_\nu ^b(y_j){\bf \psi }%
_\alpha (x_1){\bf \psi }_\beta ^c(x_2)\overline{{\bf \psi }}_\rho (y_1)%
\overline{{\bf \psi }}_\sigma ^c(y_2)]\mid 0^{-}\rangle .$%
\end{tabular}
\eqnum{ 19}
\end{equation}

In the next section, it will be shown that the Green's functions $G_\mu
^a(x_1\mid x_1,u_2;y_1,y_2)$, $G_\nu ^b(x_2\mid u_1,x_2;y_1,y_2)$ and $%
G_{\mu \nu }^{ab}(x_1,x_2\mid x_1,x_2;y_1,y_2$ are all B-S (two-particle)
reducible. Therefore, we can write 
\begin{equation}
\begin{array}{c}
G_\mu ^a(x_1\mid x_1,u_2;y_1,y_2)_{\gamma \lambda \rho \sigma }=\int
d^4z_1d^4z_2K_\mu ^{(1)a}(x_1,u_2;z_1,z_2)_{\gamma \lambda \delta \tau
}G(z_1,z_2;y_1,y_2)_{\delta \tau \rho \sigma }, \\ 
G_\nu ^b(x_2\mid u_1,x_2;y_1,y_2)_{\gamma \lambda \rho \sigma }=\int
d^4z_1d^4z_2K_\nu ^{(2)b}(u_1,x_2;z_1,z_2)_{\gamma \lambda \delta \tau
}G(z_1,z_2;y_1,y_2)_{\delta \tau \rho \sigma }, \\ 
G_{\mu \nu }^{ab}(x_1,x_2\mid x_1,x_2;y_1,y_2)_{\gamma \lambda \rho \sigma
}=\int d^4z_1d^4z_2K_{\mu \nu }^{ab}(x_1,x_2;z_1,z_2)_{\gamma \lambda \delta
\tau }G(z_1,z_2;y_1,y_2)_{\delta \tau \rho \sigma }.
\end{array}
\eqnum{20}
\end{equation}
Substituting Eq. (20) into Eq. (18) and defining 
\begin{equation}
\begin{array}{c}
K_1(x_1,u_2;z_1,z_2)_{\alpha \lambda \delta \tau }=(\Omega ^{a\mu })_{\alpha
\gamma }K_\mu ^{(1)a}(x_1,u_2;z_1,z_2)_{\gamma \lambda \delta \tau } \\ 
K_2(u_1,x_2;z_1,z_2)_{\gamma \beta \delta \tau }=(\overline{\Omega }^{b\nu
})_{\beta \lambda }K_\nu ^{(2)b}(u_1,x_2;z_1,z_2)_{\gamma \lambda \delta
\tau } \\ 
\overline{K}(x_1,x_2;z_1,z_2)_{_{\alpha \beta \delta \tau }}=(\Omega ^{a\mu
})_{\alpha \gamma }(\overline{\Omega }^{b\nu })_{\beta \lambda }K_{\mu \nu
}^{ab}(x_1,x_2;z_1,z_2)_{\gamma \lambda \delta \tau } \\ 
\overline{K}_0(x_1,x_2;z_1,z_2)_{_{\alpha \beta \delta \tau }}=\Sigma
(x_1,z_1)_{\alpha \delta }\Sigma ^c(x_2,z_2)_{\beta \tau },
\end{array}
\eqnum{21}
\end{equation}
Eq. (18) will be written as a closed form 
\begin{equation}
\begin{tabular}{l}
$\lbrack (i{\bf \partial }_{x_1}-m_1+\Sigma )(i{\bf \partial }%
_{x_2}-m_2+\Sigma ^c)G]_{\alpha \beta \rho \sigma }(x_{1,}x_2;y_1,y_2)$ \\ 
$=\delta _{\alpha \rho }\delta _{\beta \sigma }\delta ^4(x_1-y_1)\delta
^4(x_2-y_2)$ \\ 
$+\int d^4z_1d^4z_2K(x_1,x_2;z_1,z_2)_{\alpha \beta \delta \tau
}G(z_1,z_2;y_1,y_2)_{\delta \tau \rho \sigma }$%
\end{tabular}
\eqnum{22}
\end{equation}
where 
\begin{equation}
\begin{array}{c}
K(x_1,x_2;z_1,z_2)_{\alpha \beta \delta \tau }=\overline{K}%
(x_1,x_2;z_1,z_2)_{\alpha \beta \delta \tau }-\overline{K}%
_1(x_1,x_2;z_1,z_2)_{\alpha \beta \delta \tau } \\ 
-\overline{K}_2(x_1,x_2;z_1,z_2)_{\alpha \beta \delta \tau }+\overline{K}%
_0(x_1,x_2;z_1,z_2)_{_{\alpha \beta \delta \tau }}
\end{array}
\eqnum{23}
\end{equation}
is just the B-S interaction kernel in which 
\begin{equation}
\begin{array}{c}
\overline{K}_1(x_1,x_2;z_1,z_2)_{\alpha \beta \delta \tau }=\int
d^4u_2\Sigma ^c(x_2,u_2)_{\beta \lambda }K_1(x_1,u_2;y_1,y_2)_{\alpha
\lambda \delta \tau }, \\ 
\overline{K}_2(x_1,x_2;z_1,z_2)_{\alpha \beta \delta \tau }=\int
d^4u_1\Sigma (x_1,u_1)_{\alpha \gamma }K_2(u_1,x_2;y_1,y_2)_{\gamma \beta
\delta \tau }.
\end{array}
\eqnum{24}
\end{equation}

By making use of the Lehmann representation of the Green's function $%
G(x_1,x_2;y_1,y_2)$ or the well-known procedure proposed by Gell-Mann and
Low [2], one may readily derive from Eq. (22) the B-S equation satisfied by
B-S amplitudes describing the $q\overline{q\text{ }}$ bound states 
\begin{equation}
\lbrack (i{\bf \partial }_{x_1}-m_1+\Sigma )(i{\bf \partial }%
_{x_2}-m_2+\Sigma ^c)\chi _{P\varsigma }](x_{1,}x_2)=\int
d^4y_1d^4y_2K(x_1,x_2;y_1,y_2)\chi _{P\varsigma }(y_1,y_2)  \eqnum{25}
\end{equation}
where

\begin{equation}
\chi _{P\varsigma }(x_{1,}x_2)=\langle 0^{+}\mid T[{\bf \psi }(x_1){\bf \psi 
}^c(x_2)]\mid P\varsigma \rangle  \eqnum{26}
\end{equation}
represents the B-S amplitude in which $P$ denotes the total momentum of a $q%
\overline{q\text{ }}$ bound state and$\ \varsigma $ marks the other quantum
numbers of the state. The above equation can be written in the form of an
integral equation if we operating on the both sides of the above equation
with the inverse $(i{\bf \partial }_{x_1}-m_1+\Sigma )^{-1}(i{\bf \partial }%
_{x_2}-m_2+\Sigma ^c)^{-1}$%
\begin{equation}
{\cal \chi }_{P\varsigma }(x_{1,}x_2)=\int
d^4z_1d^4z_2d^4y_1d^4y_2S_F(x_1-z_1)S_F^c(x_2-z_2)K(z_1,z_2;y_1,y_2)\chi
_{P\varsigma }(y_1,y_2).  \eqnum{27}
\end{equation}

\section{B-S reducibility of the Green's functions}

The aim of this section is to analyze the B-S reducibility of the Green's
functions $G_\mu ^a(x_1\mid x_1,u_2;y_1,y_2)$, $G_\nu ^b(x_2\mid
u_1,x_2;y_1,y_2)$ and $G_{\mu \nu }^{ab}(x_1,x_2\mid x_1,x_2;y_1,y_2)$ shown
in Eq. (20). First, we start from the relation between the full $q\overline{q%
}$ four-point Green's function $G(x_{1,}x_2;y_1,y_2)$ and its connected one $%
G_c(x_{1,}x_2;y_1,y_2)$. In the case that the quark and the antiquark have
different flavors, as derived in the beginning of Appendix, this relation is
[15-17] 
\begin{equation}
G(x_{1,}x_2;y_1,y_2)=G_c(x_{1,}x_2;y_1,y_2)+S_F(x_1-y_1)S_F^c(x_2-y_2) 
\eqnum{28}
\end{equation}
Here it is noted that since the flavors of the quark and the antiquark are
different, the contraction between the quark and the antiquark vanishes,
implying that the terms related to the quark-antiquark annihilation are
absent in the above decomposition. A similar relation for the Green's
function $G_\mu ^a(x_i\mid x_1,x_2;y_1,y_2)$ can be written from Eq. (A7)
given in Appendix by setting the source $J=0$ [15-17], 
\begin{equation}
\begin{tabular}{l}
$G_\mu ^a(x_i\mid x_{1,}x_2;y_1,y_2)=G_{c\mu }^a(x_i\mid
x_{1,}x_2;y_1,y_2)+\Lambda _\mu ^a(x_i\mid x_1;y_1)S_F^c(x_2-y_2)$ \\ 
$+S_F(x_1-y_1)\Lambda _\mu ^{{\bf c}a}(x_i\mid x_2;y_2)$%
\end{tabular}
\eqnum{29}
\end{equation}
where $i=1,2$, $G_{c\mu }^a(x_i\mid x_{1,}x_2;y_1,y_2)$ is the connected
part of the Green's function $G_\mu ^a(x_i\mid x_{1,}x_2;y_1,y_2)$ and $%
\Lambda _\mu ^a(x_i\mid x_1;y_1)$ and $\Lambda _\mu ^{{\bf c}a}(x_i\mid
x_2;y_2)$ are the three-point Green's functions as defined in Eq. (16).

Let us analyze the connected Green's functions on the right-hand side of Eq.
(30) through the technique of one-particle-irreducible decomposition of
connected Green's functions. The decompositions have been carried out in the
Appendix. According to the decomposition in Eq. (A15), the three-point
gluon-quark Green's functions $\Lambda _\mu ^a(x_i\mid x_j;y_k)$ and $%
\Lambda _\mu ^{{\bf c}a}(x_i\mid x_j;y_k)$ which are fully connected can be
represented in the form 
\begin{equation}
\Lambda _\mu ^a(x_i\mid x_j;y_k)=\frac \delta {i\delta J^{a\mu }(x_i)}%
S_F(x_j-y_k)^J\mid _{J=0}=\int d^4z_1\Sigma _\mu ^a(x_i\mid
x_j;z_1)S_F(z_1-y_k)  \eqnum{30}
\end{equation}
where 
\begin{equation}
\Sigma _\mu ^a(x_i\mid x_j;z_1)=\int d^4u_1d^4u_2\Delta _{\mu \nu
}^{ab}(x_i-u_1)S_F(x_j-u_2)\Gamma ^{b\nu }(u_1\mid u_2,z_1)  \eqnum{31}
\end{equation}
and 
\begin{equation}
\Lambda _\mu ^{{\bf c}a}(x_i\mid x_j;y_k)=\frac \delta {i\delta J^{a\mu
}(x_i)}S_F^c(x_j-y_k)^J\mid _{J=0}=\int d^4z_2\Sigma _\mu ^{{\bf c}%
a}(x_i\mid x_j;z_2)S_F^c(z_2-y_k)  \eqnum{32}
\end{equation}
where 
\begin{equation}
\Sigma _\mu ^{ca}(x_i\mid x_j;z_2)=\int d^4u_1d^4u_2\Delta _{\mu \nu
}^{ab}(x_i-u_1)S_F^c(x_j-u_2)\Gamma _c^{b\nu }(u_1\mid u_2,z_2)  \eqnum{33}
\end{equation}
In the above, 
\begin{equation}
\Delta _{\mu \nu }^{ab}(x_i-u_j)=\frac 1i\left\langle 0^{+}\left| T[{\bf A}%
_\mu ^a(x_i){\bf A}_\nu ^b(u_j)]\right| 0^{-}\right\rangle =\frac 1iD_{\mu
\nu }^{ab}(x_i-u_j)  \eqnum{34}
\end{equation}
is the exact gluon propagator and $\Gamma ^{b\nu }(u_1\mid u_2,z_1)$ and $%
\Gamma _c^{b\nu }(u_1\mid u_2,z_2)$ are the gluon-quark and gluon-antiquark
three-line proper vertices respectively as defined in Eqs. (A17) and (A18).
In the case of $i=j$, the functions in Eqs. (31) and (33) give the quark and
antiquark self-energies as shown in Eq. (15). When $i\neq j$, the functions
in Eqs. (31) and (33) are related to the one gluon exchange interactions.
With the expressions in Eqs. (31) and (33), the last two terms in Eq. (29)
can be represented as 
\begin{equation}
\begin{tabular}{l}
$\Lambda _\mu ^a(x_i\mid x_1;y_1)S_F^c(x_2-y_2)+\Lambda _\mu ^{{\bf c}%
a}(x_i\mid x_2;y_2)S_F(x_1-y_1)$ \\ 
$=\int d^4z_1d^4z_2K_{0\mu
}^{(i)a}(x_1,x_2;z_1,z_2)S_F(z_1-y_1)S_F^c(z_2-y_2)$%
\end{tabular}
\eqnum{35}
\end{equation}
where 
\begin{equation}
K_{0\mu }^{(i)a}(x_1,x_2;z_1,z_2)=\Sigma _\mu ^a(x_i\mid x_1;z_1)\delta
^4(x_2-z_2)+\Sigma _\mu ^{{\bf c}a}(x_i\mid x_2;z_2)\delta ^4(x_1-z_1) 
\eqnum{36}
\end{equation}

Now we turn to the irreducible decomposition of the first term in Eq. (29).
As stated in Appendix, this decomposition may be derived from the functional
differential of the Green's function $G_c(x_{1,}x_2;y_1,y_2)$ with respect
to the source $J^{a\mu }(x_i)$ by using the one-particle irreducible
decomposition of the function $G_c(x_{1,}x_2;y_1,y_2)$. The latter
decomposition whose derivation is sketched in Appendix is well-known [15-17]
and can be represented in the form 
\begin{equation}
\begin{tabular}{l}
$G_c(x_{1,}x_2;y_1,y_2)=\int
\prod\limits_{i=1}^2d^4u_id^4v_iS_F(x_1-u_1)S_F^c(x_2-u_2)$ \\ 
$\times \Gamma (u_1,u_2;v_1,v_2)S_F(v_1-y_1)S_F^c(v_2-y_2)$%
\end{tabular}
\eqnum{37}
\end{equation}
where 
\begin{equation}
\Gamma (u_1,u_2;v_1,v_2)=\Gamma _1(u_1,u_2;v_1,v_2)+\Gamma
_2(u_1,u_2;v_1,v_2)  \eqnum{38}
\end{equation}
in which 
\begin{equation}
\Gamma _1(u_1,u_2;v_1,v_2)=-\int d^4z_1d^4z_2\Gamma ^{b\nu }(z_1\mid
u_1,v_1)D_{\nu \nu ^{\prime }}^{bb^{\prime }}(z_1-z_2)\Gamma _c^{b^{\prime
}\nu ^{\prime }}(z_2\mid u_2,v_2)  \eqnum{39}
\end{equation}
and $\Gamma _2(u_1,u_2;v_1,v_2)$ defined in Eq. (A20) is the quark-antiquark
four-line proper vertex. After substituting the expressions in Eqs.
(37)-(39), which are now given in the presence of source $J$, into Eq. (A8)
and completing the differentiation by using the differentials denoted in
Eqs. (A21) and (A23), the one-particle irreducible decomposition of the
Green's function $G_{c\mu }^a(x_i\mid x_{1,}x_2;y_1,y_2)$ will be found and,
thereby, we can write 
\begin{equation}
\begin{array}{c}
G_{c\mu }^a(x_i\mid x_{1,}x_2;y_1,y_2)=\int d^4z_1d^4z_2K_{0\mu
}^{(i)a}(x_1,x_2;z_1,z_2)G_c(z_{1,}z_2;y_1,y_2) \\ 
+\int d^4z_1d^4z_2[K_{1\mu }^{(i)a}(x_1,x_2;z_1,z_2)+K_{2\mu
}^{(i)a}(x_1,x_2;z_1,z_2)]S_F(z_1-y_1)S_F^c(z_2-y_2) \\ 
+G_{c\mu }^a(x_i\mid x_{1,}x_2;y_1,y_2)_{RE}
\end{array}
\eqnum{40}
\end{equation}
where $K_{0\mu }^{(i)a}(x_1,x_2;z_1,z_2)$ was defined in Eq. (36), 
\begin{equation}
\begin{tabular}{l}
$K_{1\mu }^{(i)a}(x_1,x_2;z_1,z_2)=\int
\prod\limits_{j=1}^2d^4u_jd^4vS_F(x_1-u_1)S_F^c(x_2-u_2)$ \\ 
$\times [\Gamma (u_1,u_2;v,z_2)\Sigma _\mu ^a(x_i\mid v;z_1)+\Gamma
(u_1,u_2;z_1,v)\Sigma _\mu ^{{\bf c}a}(x_i\mid v;z_2)],$%
\end{tabular}
\eqnum{41}
\end{equation}
\begin{equation}
K_{2\mu }^{(i)a}(x_1,x_2;z_1,z_2)=\int
\prod_{j=1}^2d^4u_jS_F(x_1-u_1)S_F^c(x_2-u_2)\overline{\Gamma }_{IR}^{a\mu
}(x_i\mid u_1,u_2;z_1,z_2)  \eqnum{42}
\end{equation}
and 
\begin{equation}
\begin{tabular}{l}
$G_{c\mu }^a(x_i\mid x_{1,}x_2;y_1,y_2)_{RE}=\int
\prod\limits_{j=1}^2d^4u_jd^4v_jd^4uS_F(x_1-u_1)S_F^c(x_2-u_2)$ \\ 
$\times \overline{\Gamma }_{RE}^{a\mu }(x_i\mid
u_1,u_2;v_1,v_2)S_F(v_1-y_1)S_F^c(v_2-y_2)$%
\end{tabular}
\eqnum{43}
\end{equation}
where $\overline{\Gamma }_{IR}^{a\mu }(x_i\mid u_1,u_2;v_1,v_2)$ and $%
\overline{\Gamma }_{RE}^{a\mu }(x_i\mid u_1,u_2;v_1,v_2)$ are respectively
the B-S irreducible and reducible parts of the vertex $\overline{\Gamma }%
^{a\mu }(x_i\mid u_1,u_2;v_1,v_2)$ defined in Eq. (A25). Here it has been
considered that the vertex $\overline{\Gamma }^{a\mu }(x_i\mid
u_1,u_2;v_1,v_2)$ is two-particle reducible (or say, B-S reducible) although
it is one-particle irreducible. This vertex is specified in the following.
Corresponding to Eq. (38), we have 
\begin{equation}
\overline{\Gamma }^{a\mu }(x_i\mid u_1,u_2;v_1,v_2)=\overline{\Gamma }%
_1^{a\mu }(x_i\mid u_1,u_2;v_1,v_2)+\overline{\Gamma }_2^{a\mu }(x_i\mid
u_1,u_2;v_1,v_2)  \eqnum{44}
\end{equation}
where 
\begin{equation}
\overline{\Gamma }_1^{a\mu }(x_i\mid u_1,u_2;v_1,v_2)=\int d^4zD_{ab}^{\mu
\nu }(x_i-z)\Gamma _{1\nu }^b(z\mid u_1,u_2;v_1,v_2)  \eqnum{45}
\end{equation}
in which 
\begin{equation}
\begin{tabular}{l}
$\Gamma _{1\nu }^b(z\mid u_1,u_2;v_1,v_2)$ \\ 
$=-\int d^4z_1d^4z_2\{\Gamma _\nu ^{b,c\rho }(z,z_1\mid u_1,v_1)D_{\rho
\sigma }^{cd}(z_1-z_2)\Gamma _{{\bf c}}^{d\sigma }(z_2\mid u_2,v_2)$ \\ 
$+\Gamma ^{c\rho }(z_1\mid u_1,v_1)[D_{\rho \sigma }^{cd}(z_1-z_2)\Gamma _{%
{\bf c}\nu }^{b,d\sigma }(z,z_2\mid u_2,v_2)+\Pi _{\nu \rho \sigma
}^{bcd}(z,z_1,z_2)\Gamma ^{d\sigma }(z_2\mid u_2,v_2)]\}$%
\end{tabular}
\eqnum{46}
\end{equation}
with 
\begin{equation}
\Pi _{\nu \rho \sigma }^{bcd}(z,z_1,z_2)=\int d^4v_1d^4v_2D_{\rho \rho
^{\prime }}^{cc^{\prime }}(z_1-v_1)\Gamma _{\nu c^{\prime }d^{\prime
}}^{b\rho ^{\prime }\sigma ^{\prime }}(z,v_1,v_2)D_{\sigma ^{\prime }\sigma
}^{d^{\prime }d}(v_2-z_2)  \eqnum{47}
\end{equation}
and $\overline{\Gamma }_2^{a\mu }(x_i\mid u_1,u_2;v_1,v_2)$ is the vertex
obtained from $\Gamma _2(u_1,u_2;v_1,v_2)$ by the differential with respect
to $J_\mu ^a(x_i)$ as defined in Eq. (A25) and represented in the following 
\begin{equation}
\overline{\Gamma }_2^{a\mu }(x_i\mid u_1,u_2;v_1,v_2)=\int d^4zD_{\mu \nu
}^{ab}(x_i-z)\Gamma _2^{b\nu }(z\mid u_1,u_2;v_1,v_2)  \eqnum{48}
\end{equation}
where $\Gamma _2^{b\nu }(z\mid u_1,u_2;v_1,v_2)$ just is the vertex $\Gamma
^{b\nu }(z\mid u_1,u_2;v_1,v_2)$ defined in Eq. (A29). It is noted that the
function $\overline{\Gamma }_1^{a\mu }(u\mid u_1,u_2;v_1,v_2)$ is B-S
irreducible as seen from its Feynman diagram; but, the vertex $\overline{%
\Gamma }_2^{a\mu }(u\mid u_1,u_2;v_1,v_2)$ is B-S reducible and can be split
into a B-S reducible part $\overline{\Gamma }_{2RE}^{a\mu }(u\mid
u_1,u_2;v_1,v_2)$ and a B-S irreducible part $\overline{\Gamma }_{2IR}^{a\mu
}(x_i\mid u_1,u_2;v_1,v_2)$. Therefore, 
\begin{equation}
\begin{array}{c}
\overline{\Gamma }_{IR}^{a\mu }(u\mid u_1,u_2;v_1,v_2)=\overline{\Gamma }%
_1^{a\mu }(u\mid u_1,u_2;v_1,v_2)+\overline{\Gamma }_{2IR}^{a\mu }(u\mid
u_1,u_2;v_1,v_2), \\ 
\overline{\Gamma }_{RE}^{a\mu }(u\mid u_1,u_2;v_1,v_2)=\overline{\Gamma }%
_{2RE}^{a\mu }(u\mid u_1,u_2;v_1,v_2).
\end{array}
\eqnum{49}
\end{equation}
Since the vertex $\overline{\Gamma }_{RE}^{a\mu }(x_i\mid u_1,u_2;v_1,v_2)$
is B-S reducible, the function $G_{c\mu }^a(x_i\mid x_{1,}x_2;y_1,y_2)_{RE}$%
, as a part of the connected Green's function, must be represented in a B-S
reducible form such that 
\begin{equation}
G_{c\mu }^a(x_i\mid x_{1,}x_2;y_1,y_2)_{RE}=\int d^4z_1d^4z_2\widetilde{K}%
_\mu ^{(i)a}(x_1,x_2;z_1,z_2)G_c(z_{1,}z_2;y_1,y_2).  \eqnum{50}
\end{equation}
As argued in Ref. [15], the kernel $\widetilde{K}_\mu
^{(i)a}(x_1,x_2;z_1,z_2)$ must be of the form 
\begin{equation}
\widetilde{K}_\mu ^{(i)a}(x_1,x_2;z_1,z_2)=K_{1\mu
}^{(i)a}(x_1,x_2;z_1,z_2)+K_{2\mu }^{(i)a}(x_1,x_2;z_1,z_2)  \eqnum{51}
\end{equation}
so as to make the B-S equation to be closed. With the above expression, when
substituting Eq. (50) into Eq. (40) and then summing up the both expressions
in Eqs. (35) and (40), noting Eq. (28), we obtain 
\begin{equation}
G_\mu ^a(x_i\mid x_1,x_2;y_1,y_2)=\int d^4z_1d^4z_2K_\mu
^{(i)a}(x_1,x_2;z_1,z_2)G(z_{1,}z_2;y_1,y_2)  \eqnum{52}
\end{equation}
where 
\begin{equation}
K_\mu ^{(i)a}(x_1,x_2;z_1,z_2)=K_{0\mu }^{(i)a}(x_1,x_2;z_1,z_2)+K_{1\mu
}^{(i)a}(x_1,x_2;z_1,z_2)+K_{2\mu }^{(i)a}(x_1,x_2;z_1,z_2).  \eqnum{53}
\end{equation}
Eq. (52) with $i=1,2$ just gives the first two equalities in Eq. (20).

Now we proceed to discuss the B-S reducibility of the Green's function $%
G_{\mu \nu }^{ab}(x_1,x_2\mid x_1,x_2;y_1,y_2)$. The decomposition of this
Green's function into the connected ones can be given by the following
calculation [16,17] 
\begin{equation}
G_{\mu \nu }^{ab}(x_1,x_2\mid x_1,x_2;y_1,y_2)=D_{\mu \nu
}^{ab}(x_1-x_2)G(x_{1,}x_2;y_1,y_2)+\frac \delta {i\delta J^{b\nu }(x_2)}%
G_\mu ^a(x_1\mid x_1,x_2;y_1,y_2)^J\mid _{J=0}  \eqnum{54}
\end{equation}
where $D_{\mu \nu }^{ab}(x_1-x_2)$ is the exact gluon propagator defined in
Eq. (34) and $G_\mu ^a(x_1\mid x_1,x_2;y_1,y_2)^J$ is the Green's function
defined in Eq. (13) in presence of the external source $J$. It is emphasized
that the expression of the $G_\mu ^a(x_1\mid x_1,x_2;y_1,y_2)^J$ formally is
the same as given in Eq. (29) or (52). On inserting Eq. (52) into Eq. (54),
completing the differentiation and using Eq. (52) once again, one can get 
\begin{equation}
G_{\mu \nu }^{ab}(x_1,x_2\mid x_1,x_2;y_1,y_2)=\int d^4z_1d^4zK_{\mu \nu
}^{ab}(x_1,x_2;z_1,z_2)G(z_{1,}z_2;y_1,y_2)  \eqnum{55}
\end{equation}
where 
\begin{equation}
K_{\mu \nu }^{ab}(x_1,x_2;z_1,z_2)=\sum\limits_{i=0}^2K_{\mu \nu
}^{(i)ab}(x_1,x_2;z_1,z_2)  \eqnum{56}
\end{equation}
with the definitions: 
\begin{equation}
K_{\mu \nu }^{(0)ab}(x_1,x_2;z_1,z_2)=D_{\mu \nu }^{ab}(x_1-x_2)\delta
^4(x_1-z_1)\delta ^4(x_2-z_2)  \eqnum{57}
\end{equation}
\begin{equation}
K_{\mu \nu }^{(1)ab}(x_1,x_2;z_1,z_2)=\frac \delta {i\delta J^{b\nu }(x_2)}%
K_\mu ^{(1)a}(x_1,x_2;z_1,z_2)^J\mid _{J=0}  \eqnum{58}
\end{equation}
which will be derived specifically in the next section and 
\begin{equation}
K_{\mu \nu }^{(2)ab}(x_1,x_2;z_1,z_2)=\int d^4u_1d^4u_2K_\mu
^{(1)a}(x_1,x_2;u_1,u_2)K_\nu ^{(2)b}(u_1,u_2;z_1,z_2)  \eqnum{59}
\end{equation}
here $K_\mu ^{(1)a}(x_1,x_2;u_1,u_2)$ and $K_\nu ^{(2)b}(u_1,u_2;z_1,z_2)$
have been expressed in Eqs. (53), (36), (41) and (42). Eq. (55) precisely
represents the B-S reducibility of the Green's function $G_{\mu \nu
}^{ab}(x_1,x_2\mid x_1,x_2;y_1,y_2)$.

\section{Expression of the B-S kernel}

Substituting Eq. (53) with the concrete expressions given in Eqs. (36), (41)
and (42) into the first two equalities in Eq. (21) and then inserting the $%
K_1(x_1,u_2;z_1,z_2)$ and $K_1(u_1,x_2;z_1,z_2)$ thus obtained into Eq.
(24), we directly get the expressions of the kernels $\overline{K}%
_1(x_1,x_2;z_1,z_2)$ and $\overline{K}_2(x_1,x_2;z_1,z_2)$. As seen from Eq.
(24) and the last equality in Eq. (21), the kernels $\overline{K}%
_1(x_1,x_2;z_1,z_2)$, $\overline{K}_2(x_1,x_2;z_1,z_2)$ and $\overline{K}%
_0(x_1,x_2;z_1,z_2)$ are all related to the self-energies of the quark and
antiquark appearing in the B-S amplitude. It will be seen that these kernels
play the role of cancelling the corresponding terms contained in the kernel $%
\overline{K}(x_1,x_2;z_1,z_2)$ defined in the third equality of Eq. (21). In
view of Eqs. (56)-(59), the kernel $\overline{K}(x_1,x_2;z_1,z_2)$ can be
written as 
\begin{equation}
\overline{K}(x_1,x_2;z_1,z_2)_{\alpha \beta \rho \sigma
}=K^{(0)}(x_1,x_2;z_1,z_2)_{\alpha \beta \rho \sigma }+\widetilde{K}%
^{(1)}(x_1,x_2;z_1,z_2)_{\alpha \beta \rho \sigma }+\widetilde{K}%
^{(2)}(x_1,x_2;z_1,z_2)_{\alpha \beta \rho \sigma }  \eqnum{60}
\end{equation}
where 
\begin{equation}
K^{(0)}(x_1,x_2;z_1,z_2)=(\Omega ^{a\mu })_{\alpha \rho }(\overline{\Omega }%
^{b\nu })_{\beta \sigma }D_{\mu \nu }^{ab}(x_1-x_2)\delta ^4(x_1-z_1)\delta
^4(x_2-z_2)  \eqnum{61}
\end{equation}
is the one-gluon exchange kernel for the t-channel interaction, 
\begin{equation}
\widetilde{K}^{(1)}(x_1,x_2;z_1,z_2)_{\alpha \beta \rho \sigma }==(\Omega
^{a\mu })_{\alpha \gamma }(\overline{\Omega }^{b\nu })_{\beta \lambda
}K_{\mu \nu }^{(1)ab}(x_1,x_2;z_1,z_2)_{\gamma \lambda \rho \sigma } 
\eqnum{62}
\end{equation}
and 
\begin{equation}
\widetilde{K}^{(2)}(x_1,x_2;z_1,z_2)_{\alpha \beta \rho \sigma }==(\Omega
^{a\mu })_{\alpha \gamma }(\overline{\Omega }^{b\nu })_{\beta \lambda
}K_{\mu \nu }^{(2)ab}(x_1,x_2;z_1,z_2)_{\gamma \lambda \rho \sigma } 
\eqnum{63}
\end{equation}

To prove the aforementioned cancellation, let us compute the functions $%
K_{\mu \nu }^{(1)ab}(x_1,x_2;z_1,z_2)$ and $K_{\mu \nu
}^{(2)ab}(x_1,x_2;z_1,z_2)$ defined in Eqs. (58) and (59). According to the
expressions in Eqs. (53) and (36), it is convenient to write 
\begin{equation}
K_\mu ^{(1)a}(x_1,x_2;z_1,z_2)=\Sigma _\mu ^a(x_1\mid x_1;z_1)\delta
^4(x_2-z_2)+\widehat{K}_\mu ^{(1)a}(x_1,x_2;z_1,z_2)  \eqnum{64}
\end{equation}
where 
\begin{equation}
\widehat{K}_\mu ^{(1)a}(x_1,x_2;z_1,z_2)=\Sigma _\mu ^{{\bf c}a}(x_1\mid
x_2;z_2)\delta ^4(x_1-z_1)+K_{1\mu }^{(1)a}(x_1,x_2;z_1,z_2)+K_{2\mu
}^{(1)a}(x_1,x_2;z_1,z_2)  \eqnum{65}
\end{equation}
When Eq. (64) is inserted into Eq. (59) and then Eq. (59) is inserted into
Eq. (63), considering the second equality in Eq. (24), we get 
\begin{equation}
\begin{array}{c}
\widetilde{K}^{(2)}(x_1,x_2;z_1,z_2)_{_{\alpha \beta \rho \sigma }}=%
\overline{K}_2(x_1,x_2;z_1,z_2)_{\alpha \beta \rho \sigma } \\ 
+\int d^4u_1d^4u_2(\Omega ^{a\mu })_{\alpha \gamma }(\overline{\Omega }%
^{b\nu })_{\beta \lambda }\widehat{K}_\mu ^{(1)a}(x_1,x_2;u_1,u_2)_{\gamma
\delta \lambda \tau }K_\nu ^{(2)b}(u_1,u_2;z_1,z_2)_{\lambda \tau \rho
\sigma }
\end{array}
\eqnum{66}
\end{equation}
In the above, we have considered that $\Sigma (x_1,z_1)_{\alpha \rho
}=(\Omega ^{a\mu })_{\alpha \gamma }\Sigma _\mu ^a(x_1\mid x_1;z_1)_{\gamma
\rho }$ is the quark self-energy.

Now, let us turn to compute $K_{\mu \nu }^{(1)ab}(x_1,x_2;z_1,z_2)$. Looking
at the expressions in Eqs. (33), (41) and (42), it is seen that the function 
$\widehat{K}_\mu ^{(1)a}(x_1,x_2;z_1,z_2)$ can be represented in the form 
\begin{equation}
\widehat{K}_\mu ^{(1)a}(x_1,x_2;z_1,z_2)=\int d^4u_2S_F^c(x_2-u_2)Q_\mu
^a(x_1\mid x_1,u_2;z_1,z_2)  \eqnum{67}
\end{equation}
where 
\begin{equation}
\begin{array}{c}
Q_\mu ^a(x_1\mid x_1,u_2;z_1,z_2)=Q_\mu ^{(0)a}(x_1\mid x_1,u_2;z_1,z_2) \\ 
+Q_\mu ^{(1)a}(x_1\mid x_1,u_2;z_1,z_2)+Q_\mu ^{(2)a}(x_1\mid
x_1,u_2;z_1,z_2)
\end{array}
\eqnum{68}
\end{equation}
in which 
\begin{equation}
Q_\mu ^{(0)a}(x_1\mid x_1,u_2;z_1,z_2)=\int d^4u_1\Delta _{\mu \nu
}^{ab}(x_1-u_1)\Gamma ^{b\nu }(u_1\mid u_2,z_2)\delta ^4(x_1-z_1)  \eqnum{69}
\end{equation}
\begin{equation}
\begin{array}{c}
Q_\mu ^{(1)a}(x_1\mid x_1,u_2;z_1,z_2)=\int d^4u_1d^4uS_F(x_1-u_1)[\Gamma
(u_1,u_2;u,z_2) \\ 
\times \Sigma _\mu ^a(x_1\mid u,z_1)+\Gamma (u_1,u_2;z_1,u)\Sigma _\mu ^{%
{\bf c}a}(x_1\mid u,z_2)]
\end{array}
\eqnum{70}
\end{equation}
and 
\begin{equation}
Q_\mu ^{(2)a}(x_1\mid x_1,u_2;z_1,z_2)=\int d^4u_1S_F(x_1-u_1))\overline{%
\Gamma }_{IR}^{b\nu }(x_1\mid u_1,u_2;z_1,z_2).  \eqnum{71}
\end{equation}
When the expression in Eq. (64) with the expression in Eq. (67) is inserted
into Eq. (58) and completing the differentiation, noticing Eq. (32), one can
find 
\begin{equation}
\begin{array}{c}
K_{\mu \nu }^{(1)ab}(x_1,x_2;z_1,z_2)=\Sigma _{\mu \nu }^{ab}(x_1,x_2\mid
x_1,z_1)\delta ^4(x_2-z_2) \\ 
+\int d^4v\Sigma _\nu ^{{\bf c}b}((x_2\mid x_2,v)\widehat{K}_\mu
^{(1)a}(x_1,v;z_1,z_2) \\ 
+\int d^4u_2S_F^c(x_2-u_2)Q_{\mu \nu }^{ab}(x_1,x_2\mid x_1,u_2;z_1,z_2)
\end{array}
\eqnum{72}
\end{equation}
where 
\begin{equation}
\Sigma _{\mu \nu }^{ab}(x_1,x_2\mid x_1,z_1)=\frac \delta {i\delta J^{b\nu
}(x_2)}\Sigma _\mu ^a(x_1\mid x_1,z_1)^J\mid _{J=0}  \eqnum{73}
\end{equation}
and 
\begin{equation}
Q_{\mu \nu }^{ab}(x_1,x_2\mid x_1,u_2;z_1,z_2)=\frac \delta {i\delta J^{b\nu
}(x_2)}Q_\mu ^a(x_1\mid x_1,u_2;z_1,z_2)^J\mid _{J=0}  \eqnum{74}
\end{equation}
which will be calculated in detail soon later. According to Eq. (64), the
second term in Eq. (72) can be represented as 
\begin{equation}
\begin{array}{c}
\int d^4v\Sigma _\nu ^{{\bf c}b}((x_2\mid x_2,v)\widehat{K}_\mu
^{(1)a}(x_1,v;z_1,z_2) \\ 
=\int d^4v\Sigma _\nu ^{{\bf c}b}((x_2\mid x_2,v)K_\mu
^{(1)a}(x_1,v;z_1,z_2)-\Sigma _\mu ^a(x_1\mid x_1;z_1)\Sigma _\nu ^{{\bf c}%
b}((x_2\mid x_2,z_2)
\end{array}
\eqnum{75}
\end{equation}
Noticing this expression, when Eq. (72) is substituted into Eq. (62), we
have 
\begin{equation}
\begin{array}{c}
\widetilde{K}^{(1)}(x_1,x_2;z_1,z_2)_{_{\alpha \beta \rho \sigma }}=(\Omega
^{a\mu })_{\alpha \gamma }\Sigma _{\mu \nu }^{ab}(x_1,x_2\mid
x_1,z_1)_{\gamma \lambda \rho \sigma }(\overline{\Omega }^{b\nu })_{\beta
\sigma }\delta ^4(x_2-z_2) \\ 
+\int d^4u_2(\Omega ^{a\mu })_{\alpha \gamma }(\overline{\Omega }^{b\nu
})_{\beta \lambda }S_F^c(x_2-u_2)_{\lambda \tau }Q_{\mu \nu
}^{ab}(x_1,x_2\mid x_1,u_2;z_1,z_2)_{\gamma \tau \rho \sigma } \\ 
+\overline{K}_1(x_1,x_2;z_1,z_2)_{\alpha \beta \rho \sigma }-\overline{K}%
_0(x_1,x_2;z_1,z_2)_{_{\alpha \beta \rho \sigma }}
\end{array}
\eqnum{76}
\end{equation}
where the definitions given in Eqs. (21) and (24) as well as $\Sigma
^c(x_2,z_2)_{\beta \sigma }=(\overline{\Omega }^{b\nu })_{\beta \lambda
}\Sigma _\nu ^{{\bf c}b}(x_2\mid x_2;z_2)_{\lambda \sigma }$ which is the
antiquark self-energy have been noted.

It is clear that when Eq. (60) is substituted into Eq. (23) and considering
the expressions in Eqs. (66) and (76), we see, the last three terms in Eq.
(23) are all cancelled out. As a result of the cancellation, we have 
\begin{equation}
\begin{array}{c}
K(x_1,x_2;z_1,z_2)_{\alpha \beta \rho \sigma
}=K^{(0)}(x_1,x_2;z_1,z_2)_{\alpha \beta \rho \sigma } \\ 
+(\Omega ^{a\mu })_{\alpha \gamma }\Sigma _{\mu \nu }^{ab}(x_1,x_2\mid
x_1,z_1)_{\gamma \lambda \rho \sigma }(\overline{\Omega }^{b\nu })_{\beta
\sigma }\delta ^4(x_2-z_2) \\ 
+\int d^4u_2(\Omega ^{a\mu })_{\alpha \gamma }(\overline{\Omega }^{b\nu
})_{\beta \lambda }S_F^c(x_2-u_2)_{\lambda \tau }Q_{\mu \nu
}^{ab}(x_1,x_2\mid x_1,u_2;z_1,z_2)_{\gamma \tau \rho \sigma } \\ 
+\int d^4u_1d^4u_2(\Omega ^{a\mu })_{\alpha \gamma }(\overline{\Omega }%
^{b\nu })_{\beta \lambda }\widehat{K}_\mu ^{(1)a}(x_1,x_2;u_1,u_2)_{\gamma
\delta \lambda \tau }K_\nu ^{(2)b}(u_1,u_2;z_1,z_2)_{\lambda \tau \rho
\sigma }
\end{array}
\eqnum{77}
\end{equation}
To give an explicit expression of the above kernel, we need to compute the
functions $\Sigma _{\mu \nu }^{ab}(x_1,x_2\mid x_1,z_1)$ and $Q_{\mu \nu
}^{ab}(x_1,x_2\mid x_1,u_2;z_1,z_2)$. Substituting Eq. (31) into Eq. (73)
and employing the formulas as given in Eqs. (30), (32), (A21) and (A23) to
complete the differentiation, it is easy to get 
\begin{equation}
\begin{array}{c}
\Sigma _{\mu \nu }^{ab}(x_1,x_2\mid x_1,z_1)=\int
\prod\limits_{i=1}^3d^4u_i\{\Sigma _\nu ^b(x_2\mid
x_1,u_1)S_F(u_1-u_2)\Delta _{\mu \lambda }^{ac}(x_1-u_3) \\ 
\times \Gamma ^{c\lambda }(u_3\mid u_2,z_1)+S_F(x_1-u_1)\Delta _{\nu \tau
}^{bd}(x_2-u_2)[\Pi _{d\mu \lambda }^{\tau ac}(u_2,x_1,u_3)\Gamma ^{c\lambda
}(u_3\mid u_1,z_1) \\ 
+\Delta _{\mu \lambda }^{ac}(x_1-u_3)\Gamma ^{d\tau ,c\lambda }(u_2,u_3\mid
u_1,z_1)]\}
\end{array}
\eqnum{78}
\end{equation}
where $\Pi _{d\mu \lambda }^{\tau ac}(u_2,x_2,u_3)$ was defined in Eq. (47)
and $\Gamma ^{d\tau ,c\lambda }(u_2,u_3\mid u_1,z_1)$ is a kind of
gluon-quark four-line vertex defined in Eq. (A27).

In accordance with Eq. (68), the function $Q_{\mu \nu }^{ab}(x_1,x_2\mid
x_1,u_2;z_1,z_2)$ in Eq. (77) can be written as 
\begin{equation}
\begin{array}{c}
Q_{\mu \nu }^{ab}(x_1,x_2\mid x_1,u_2;z_1,z_2)=Q_{\mu \nu
}^{(0)ab}(x_1,x_2\mid x_1,u_2;z_1,z_2) \\ 
+Q_{\mu \nu }^{(1)ab}(x_1,x_2\mid x_1,u_2;z_1,z_2)+Q_{\mu \nu
}^{(2)ab}(x_1,x_2\mid x_1,u_2;z_1,z_2)
\end{array}
\eqnum{79}
\end{equation}
The three terms on the right-hand side of Eq. (79) can easily be derived
from the expressions written in Eqs. (69)-(71) by applying the formulas
denoted in Eqs. (30), (32), (A21) and (A23). The results are displayed
below. 
\begin{equation}
\begin{array}{c}
Q_{\mu \nu }^{(0)ab}(x_1,x_2\mid x_1,u_2;z_1,z_2)=\frac \delta {i\delta
J^{b\nu }(x_2)}Q_\mu ^{(0)a}(x_1\mid x_1,u_2;z_1,z_2)^J\mid _{J=0} \\ 
=\int d^4u_1d^4u_3\Delta _{\nu \lambda }^{bc}(x_2-u_3)\{\Pi _{c\mu \tau
}^{\lambda ad}(u_3,x_1,u_1)\Gamma _{{\bf c}}^{d\tau }(u_1\mid u_2,z_2) \\ 
+D_{\mu \tau }^{ad}(x_1-u_1)\Gamma _{cd}^{\lambda \tau }(u_3,u_1\mid
u_2,z_2)\}\delta ^4(x_1-z_1),
\end{array}
\eqnum{80}
\end{equation}
\begin{equation}
\begin{array}{c}
Q_{\mu \nu }^{(1)ab}(x_1,x_2\mid x_1,u_2;z_1,z_2)=\frac \delta {i\delta
J^{b\nu }(x_2)}Q_\mu ^{(1)a}(x_1\mid x_1,u_2;z_1,z_2)^J\mid _{J=0} \\ 
=\int d^4u_1d^4u_3d^4u\{\Sigma _\nu ^b(x_2\mid x_1,u_1)S_F(u_1-u_3)[\Gamma
(u_3,u_2;u,z_2)\Sigma _\mu ^a(x_1\mid u,z_1) \\ 
+\Gamma (u_3,u_2;z_1,u)\Sigma _\mu ^{{\bf c}a}(x_1\mid
u,z_2)]+S_F(x_1-u_1)D_{\nu \lambda }^{bc}(x_2-u_3)[\widehat{\Gamma }%
^{c\lambda }(u_3\mid u_1,u_2;u,z_2) \\ 
\times \Sigma _\mu ^a(x_1\mid u,z_1)+\widehat{\Gamma }^{c\lambda }(u_3\mid
u_1,u_2;z_1,u)\Sigma _\mu ^{{\bf c}a}(x_1\mid u,z_2)]\} \\ 
+\int d^4u_1d^4uS_F(x_1-u_1)[\Gamma (u_1,u_2;u,z_2)\Sigma _{\mu \nu
}^{ab}(x_1,x_2\mid u,z_1)+\Gamma (u_1,u_2;z_1,u) \\ 
\times \Sigma _{\mu \nu }^{{\bf c}ab}(x_1,x_2\mid u,z_2)]
\end{array}
\eqnum{81}
\end{equation}
where 
\begin{equation}
\widehat{\Gamma }^{b\nu }(u\mid u_1,u_2;z_1,z_2)=\Gamma _1^{b\nu }(u\mid
u_1,u_2;z_1,z_2)+\Gamma _2^{b\nu }(u\mid u_1,u_2;z_1,z_2)  \eqnum{82}
\end{equation}
in which $\Gamma _1^{b\nu }(u\mid u_1,u_2;z_1,z_2)$ was represented in Eq.
(46) and the function $\Gamma _2^{b\nu }(u\mid u_1,u_2;z_1,z_2)$ appearing
in Eq. (48) just is the vertex $\Gamma ^{b\nu }(u\mid u_1,u_2;z_1,z_2)$
defined in Eq. (A29), 
\begin{equation}
\begin{array}{c}
\Sigma _{\mu \nu }^{{\bf c}ab}(x_1,x_2\mid u,z_2)=\int
\prod\limits_{i=1}^3d^4u_i\{\Sigma _\nu ^{{\bf c}b}(x_2\mid
u,u_1)S_F^c(u_1-u_2)\Delta _{\mu \lambda }^{ac}(x_1-u_3) \\ 
\times \Gamma _{{\bf c}}^{c\lambda }(u_3\mid u_2,z_2)+S_F^c(u-u_1)\Delta
_{\nu \tau }^{bd}(x_2-u_2)[\Pi _{d\mu \lambda }^{\tau ac}(u_2,x_1,u_3)\Gamma
^{c\lambda }(u_3\mid u_1,z_2) \\ 
+\Delta _{\mu \lambda }^{ac}(x_1-u_3)\Gamma _c^{d\tau ,c\lambda
}(u_2,u_3\mid u_1,z_2)]\}
\end{array}
\eqnum{83}
\end{equation}
and $\Sigma _{\mu \nu }^{ab}(x_1,x_2\mid u,z_1)$ was formulated in Eq. (78). 
\begin{equation}
\begin{array}{c}
Q_{\mu \nu }^{(2)ab}(x_1,x_2\mid x_1,u_2;z_1,z_2)=\frac \delta {i\delta
J^{b\nu }(x_2)}Q_\mu ^{(2)a}(x_1\mid x_1,u_2;z_1,z_2)^J\mid _{J=0} \\ 
=\int d^4u_1d^4u_3d^4u\{\Sigma _\nu ^b(x_2\mid x_1,u_3)S_F(u_3-u_1)D_{\mu
\lambda }^{ac}(x_1-u)\Gamma _{IR}^{c\lambda }(u\mid u_1,u_2;z_1,z_2) \\ 
+S_F(x_1-u_1)D_{\nu \lambda }^{bc}(x_2-u_3)[\Pi _{c\mu \tau }^{\lambda
ad}(u_3,x_1,u)\Gamma _{IR}^{d\tau }(u\mid u_1,u_2;z_1,z_2) \\ 
+D_{\mu \tau }^{ad}(x_1-u)\Gamma _{IR}^{d\tau ,c\lambda }(u,u_3\mid
u_1,u_2;z_1,z_2)]\}
\end{array}
\eqnum{84}
\end{equation}
where we have considered 
\begin{equation}
\overline{\Gamma }_{IR}^{a\mu }(x_1\mid u_1,u_2;z_1,z_2)=\int d^4uD_{\mu
\lambda }^{ac}(x_1-u)\Gamma _{IR}^{c\lambda }(u\mid u_1,u_2;z_1,z_2) 
\eqnum{85}
\end{equation}
which follows from the relations in Eqs. (45) and (48), the $\Gamma
_{IR}^{c\lambda }(u\mid u_1,u_2;z_1,z_2)$ stands for the part of the $%
\widehat{\Gamma }^{c\lambda }(u\mid u_1,u_2;z_1,z_2)$ defined in Eq. (82)
which gives the B-S irreducible part of the $\overline{\Gamma }^{a\mu
}(x_i\mid u_1,u_2;z_1,z_2)$ and $\Gamma _{IR}^{d\tau ,c\lambda }(u,u_3\mid
u_1,u_2;z_1,z_2)$ designates the part of the six-point gluon-quark-antiquark
vertex $\widehat{\Gamma }^{d\tau ,c\lambda }(u,u_3\mid u_1,u_2;z_1,z_2)$
which leads to the B-S irreducible part of the vertex $\overline{\Gamma }%
_{\mu \nu }^{ab}(x_{1,}x_2\mid u_1,u_2;z_1,z_2$) defined in Eq. (A26). By
using Eqs. (A21) and (A23) and the following differential: 
\begin{equation}
\frac \delta {i\delta J^{b\nu }(x_2)}\Pi _{\mu \rho \sigma
}^{acd}(z,z_1,z_2)^J\mid _{J=0}=\int d^4uD_{\mu \lambda }^{bc}(x_2-u)\Pi
_{c\mu \rho \sigma }^{\lambda ade}(u,z,z_1,z_2)  \eqnum{86}
\end{equation}
where 
\begin{equation}
\begin{array}{c}
\Pi _{\lambda \mu \rho \sigma }^{cade}(u,z,z_1,z_2)=\int d^4u_1d^4u_2\{\Pi
_{\lambda \rho \rho ^{\prime }}^{cdd^{\prime }}(u,z_1,u_1)\Gamma _{\mu
d^{\prime }e^{\prime }}^{a\rho ^{\prime }\sigma ^{\prime
}}(z,u_1,u_2)D_{\sigma ^{\prime }\sigma }^{e^{\prime }e}(u_2-z_2) \\ 
+D_{\rho \rho ^{\prime }}^{dd^{\prime }}(z_1-u_1)[\Gamma _{\lambda \mu
d^{\prime }e^{\prime }}^{ca\rho ^{\prime }\sigma ^{\prime
}}(u,z,u_1,u_2)D_{\sigma ^{\prime }\sigma }^{e^{\prime }e}(u_2-z_2)+\Gamma
_{\mu d^{\prime }e^{\prime }}^{a\rho ^{\prime }\sigma ^{\prime
}}(z,u_1,u_2)\Pi _{\lambda \sigma ^{\prime }\sigma }^{ce^{\prime
}e}(u,u_2,z_2)]\}
\end{array}
\eqnum{87}
\end{equation}
with $\Gamma _{\lambda \mu d^{\prime }e^{\prime }}^{ca\rho ^{\prime }\sigma
^{\prime }}(u,z,u_1,u_2)$ being the gluon four-line vertex, it is not
difficult to derive from Eqs. (82) and (46) the expression of $\widehat{%
\Gamma }_{\lambda \tau }^{cd}(u,u_3\mid u_1,u_2;z_1,z_2)$ which is shown in
the following: 
\begin{equation}
\widehat{\Gamma }_{\lambda \tau }^{cd}(u,u_3\mid u_1,u_2;z_1,z_2)=\Gamma
_{1\lambda \tau }^{cd}(u,u_3\mid u_1,u_2;z_1,z_2)+\Gamma _{2\lambda \tau
}^{cd}(u,u_3\mid u_1,u_2;z_1,z_2)  \eqnum{88}
\end{equation}
where 
\begin{equation}
\begin{array}{c}
\Gamma _{1\lambda \tau }^{cd}(u,u_3\mid u_1,u_2;z_1,z_2)=-\int
d^4v_1d^4v_2\{\Gamma _{\lambda \tau }^{cd,e\sigma }(u,u_3,v_1\mid
u_1,z_1)D_{\sigma \rho }^{ef}(v_1-v_2)\Gamma _{{\bf c}}^{f\rho }(v_2\mid
u_2,z_2) \\ 
+\Gamma _\lambda ^{c,e\sigma }(u,,v_1\mid u_1,z_1)[\Pi _{\tau \sigma \rho
}^{def}(u_3,v_1,v_2)\Gamma _{{\bf c}}^{f\rho }(v_2\mid u_2,z_2)+D_{\sigma
\rho }^{ef}(v_1-v_2)\Gamma _{{\bf c}\tau }^{d,,f\rho }(u_3,v_2\mid u_2,z_2)]
\\ 
+\Gamma _\tau ^{d,e\sigma }(u_3,v_1\mid u_1,z_1)D_{\sigma \rho
}^{ef}(v_1-v_2)\Gamma _{{\bf c}\lambda }^{c,,f\rho }(u,v_2\mid
u_2,z_2)+\Gamma ^{e\sigma }(v_1\mid u_1,z_1)[\Pi _{\tau \sigma \rho
}^{def}(u_3,v_1,v_2) \\ 
\times \Gamma _{{\bf c}}^{f\rho }(v_2\mid u_2,z_2)+D_{\sigma \rho
}^{ef}(v_1-v_2)\Gamma _{{\bf c}\lambda \tau }^{cd,,f\rho }(u,u_3,v_2\mid
u_2,z_2)]+\Gamma _\tau ^{d,e\sigma }(u_3,v_1\mid u_1,z_1)\Pi _{\lambda
\sigma \rho }^{cef}(u,v_1,v_2) \\ 
\times \Gamma _{{\bf c}}^{f\rho }(v_2\mid u_2,z_2)+\Gamma ^{e\sigma
}(v_1\mid u_1,z_1)[\Pi _{\lambda \tau \sigma \rho
}^{cdef}(u,u_3,v_1,v_2)\Gamma _{{\bf c}}^{f\rho }(v_2\mid u_2,z_2)+\Pi
_{\lambda \sigma \rho }^{cef}(u,v_1,v_2) \\ 
\times \Gamma _{{\bf c}\tau }^{d,,f\rho }(u_3,v_2\mid u_2,z_2)]\},
\end{array}
\eqnum{89}
\end{equation}
$\Gamma _{2\lambda \tau }^{cd}(u,u_3\mid u_1,u_2;z_1,z_2)$ is the six-point
vertex $\Gamma _{\lambda \tau }^{cd}(u,u_3\mid u_1,u_2;z_1,z_2)$ defined in
Eq. (A30) and the other vertices can be read from Eq. (A24).

At last, we would like to note that the B-S kernel derived in Eqs. (77)-(89)
with $\widehat{K}_\mu ^{(1)a}(x_1,x_2;u_1,u_2)$ being given in Eq. (65) and $%
K_\nu ^{(2)b}(u_1,u_2;z_1,z_2)$ in Eq. (53) with $i=2$ is symmetric with
respect to quark and antiquark. This point can clearly be seen when all the
terms in the kernel are represented by Feynman diagrams. For example, there
are three terms in the second term in Eq. (77) which are given by the three
terms in Eq. (78). The two terms given by the last two terms in Eq. (78)
have an one-to-one correspondence with the two terms appearing in the third
terms in Eq. (77) which are given by the two terms in Eq. (80). The
remaining term given by the first term in Eq. (78) corresponds to such a
terms included in the last term in Eq. (77) that it is given by the term $%
\Sigma _\mu ^{{\bf c}a}(x_1\mid x_2;u_2)\delta ^4(x_1-u_1)$ in $\widehat{K}%
_\mu ^{(1)a}(x_1,x_2;u_1,u_2)$ and the term $\Sigma _\nu ^{{\bf c}b}(x_2\mid
u_2;z_2)\delta ^4(u_1-z_1)$ in $K_\nu ^{(2)b}(u_1,u_2;z_1,z_2)$ and can be
represented as 
\begin{equation}
\int \prod\limits_{i=1}^3d^4u_i\Delta _{\mu \lambda }^{ac}(x_1-u_1)(%
\overline{\Omega }^{b\nu })_{\beta \gamma }S_F^c(x_2-u_2)_{\gamma \delta
}\Gamma _c^{c\lambda }(u_1\mid u_2,u_3)_{\delta \tau }\Sigma _\nu ^{{\bf c}%
b}(x_2\mid u_3;z_2)_{\tau \sigma }(\Omega ^{a\mu })_{\alpha \rho }\delta
^4(x_1-z_1)  \eqnum{90}
\end{equation}
The six terms mentioned above give a higher order correction to the vertices
in the one-gluon exchange kernel.

\section{Summary and remarks}

In this paper, we have derived a new expression of the B-S interaction
kernel for quark-antiquark bound states by means of the technique of
irreducible decomposition of Green's functions. The kernel given in the case
that the quark and the antiquark have different flavors was expressed in Eq.
(77). In Eq. (77), the $K^{(0)}(x_1,x_2;z_1,z_2)$ is the one-gluon exchange
interaction kernel represented in Eq. (61), the function $\Sigma _{\mu \nu
}^{ab}(x_1,x_2\mid x_1,z_1)$ was described in Eq. (78), the function $Q_{\mu
\nu }^{ab}(x_1,x_2\mid x_1,u_2;z_1,z_2)$ was formulated in detail in Eqs.
(79)-(89) and the functions $\widehat{K}_\mu ^{(1)a}(x_1,x_2;u_1,u_2)$ and $%
K_\nu ^{(2)b}(u_1,u_2;z_1,z_2)$ were specified in Eqs. (64), (65), (36),
(41) and (42). In comparison with the previous expression derived in Ref.
[14] which is compactly represented in terms of a few types of Green's
functions, expression of the kernel derived in this paper is represented in
terms of the quark, antiquark and gluon propagators and some kinds of quark,
antiquark and/or gluon three, four, five and six-point vertices and
therefore exhibits a more specific structure of kernel. It is noted that
although the kernel given in this paper is limited to the case that the
quark and the antiquark are of different flavors, it is sufficient to solve
the problem of quark confinement because the strong interaction between the
quark and antiquark which have the same flavors is the same as the one
between the quark and antiquark which are of different flavors.

In previous investigations of meson spectrum within the framework of B-S
equation [12,19], one often used the quark potential model in which besides
the one-gluon exchange kernel denoted in Eq. (61), a phenomenological
confining potential (for instance, the linear potential) is necessary to be
introduced. As we see, the confining potential is used to simulate all the
other terms included in the expression shown in Eq. (77). Obviously, the
simulation is oversimplified. In the investigations within the framework of
Dyson-Schwinger equation [20,21], since the equation is not closed,
containing an infinite set of coupled equations, in practical calculations,
one has to cut-off the equations involving higher order (more line)
vertices, and only uses the truncated equations related to the lowest order
vertices [22], it is apparent that use of the lowest order vertices is
difficult to exactly consider the effect of the terms involving higher order
vertices in Eq. (77). Since the kernel derived is exact, containing all the
interactions taking place in the bound states, clearly, one way of solving
the quark confinement is to perform a nonperturbative calculation of the
kernel presented in this paper or in our previous paper [14,15]]. Even
though the kernel given in this paper has a complicated structure,
calculation of it is feasible. We suggest that in practical calculations,
one may utilize the Ward-Takahashi identities to reduce the higher order
vertices to the lower order ones and choose a special gauge to simplify the
calculations.

\section{\bf Acknowledgment}

This project was supposed by National Natural Science Foundation of China.

\section{Appendix: One-particle irreducible decompositions of the connected
Green's functions}

Let us begin with the relation between the generating functional for full
Green's functions $Z[J,\overline{\eta },\eta ,\overline{\xi },\xi ]$ and the
one for connected Green's functions $W[J,\overline{\eta },\eta ,\overline{%
\xi },\xi ]$ [16,17] 
\begin{equation}
Z[J,\overline{\eta },\eta ,\overline{\xi },\xi ]=\exp \{iW[J,\overline{\eta }%
,\eta ,\overline{\xi },\xi ]\}  \eqnum{A1}
\end{equation}
Taking the derivatives of Eq. (A1) with respect to the sources $\overline{%
\eta }(x_1),\overline{\eta }^c(x_2),\eta (y_1)$ and $\eta ^c(y_2)$ and then
setting all the sources except for the source $J$ to be zero, in the case
that the quark and antiquark are of different flavors, one may obtain the
following decomposition 
\begin{equation}
G(x_{1,}x_2;y_1,y_2)^J=G_c(x_{1,}x_2;y_1,y_2)^J+S_F(x_1-y_1)^JS_F^c(x_2-y_2)^J
\eqnum{A2}
\end{equation}
where $G(x_{1,}x_2;y_1,y_2)^J,$ $G_c(x_{1,}x_2;y_1,y_2)^J$, $%
S_F^c(x_2-y_2)^J $ and $S_F(x_1-y_1)^J$ are defined by 
\begin{equation}
G(x_{1,}x_2;y_1,y_2)^J=\frac{\delta ^4Z[J,\overline{\eta },\eta ,\overline{%
\xi },\xi ]}{\delta \overline{\eta }(x_1)\delta \overline{\eta }%
^c(x_2)\delta \eta (y_1)\delta \eta ^c(y_2)}\mid _{\overline{\eta }=\eta =%
\overline{\xi }=\xi =0}  \eqnum{A3}
\end{equation}
\begin{equation}
G_c(x_{1,}x_2;y_1,y_2)^J=i\frac{\delta ^4W[J,\overline{\eta },\eta ,%
\overline{\xi },\xi ]}{\delta \overline{\eta }(x_1)\delta \overline{\eta }%
^c(x_2)\delta \eta (y_1)\delta \eta ^c(y_2)}\mid _{\overline{\eta }=\eta =%
\overline{\xi }=\xi =0},  \eqnum{A4}
\end{equation}
\begin{equation}
S_F(x_1-y_1)^J=\frac{\delta ^2Z[J,\overline{\eta },\eta ,\overline{\xi },\xi
]}{i\delta \overline{\eta }(x_1)\delta \eta (y_1)}\mid _{\overline{\eta }%
=\eta =\overline{\xi }=\xi =0}  \eqnum{A5}
\end{equation}
and 
\begin{equation}
S_F^c(x_2-y_2)^J=\frac{\delta ^2Z[J,\overline{\eta },\eta ,\overline{\xi }%
,\xi ]}{i\delta \overline{\eta }^c(x_2)\delta \eta ^c(y_2)}\mid _{\overline{%
\eta }=\eta =\overline{\xi }=\xi =0}  \eqnum{A6}
\end{equation}
When we set $J=0$, Eq. (A2) will go over to the decomposition shown in Eq.
(28). Differentiating Eq. (A2) with respect to the source $J^{a\mu }(x_i)$,
we have 
\begin{equation}
\begin{tabular}{l}
$G_\mu ^a(x_i\mid x_{1,}x_2;y_1,y_2)^J=G_{c\mu }^a(x_i\mid
x_{1,}x_2;y_1,y_2)^J$ \\ 
$+\Lambda _\mu ^a(x_i\mid x_1;y_1)^JS_F^c(x_2-y_2)^J+S_F(x_1-y_1)^J\Lambda
_\mu ^{{\bf c}a}(x_i\mid x_2;y_2)^J$%
\end{tabular}
\eqnum{A7}
\end{equation}
where 
\begin{equation}
G_{c\mu }^a(x_i\mid x_{1,}x_2;y_1,y_2)^J=\frac \delta {i\delta J^{a\mu }(x_i)%
}G_c(x_{1,}x_2;y_1,y_2)^J  \eqnum{A8}
\end{equation}
\begin{equation}
\Lambda _\mu ^a(x_i\mid x_1;y_1)^J=\frac \delta {i\delta J^{a\mu }(x_i)}%
S_F(x_1-y_1)^J  \eqnum{A9}
\end{equation}
\begin{equation}
\Lambda _\mu ^{{\bf c}a}(x_i\mid x_2;y_2)^J=\frac \delta {i\delta J^{a\mu
}(x_i)}S_F^c(x_2-y_2)^J  \eqnum{A10}
\end{equation}
Upon setting $J=0,$ Eq. (A7) immediately gives rise to the decomposition in
Eq. (29).

Now, let us proceed to carry out one-particle-irreducible decompositions of
the connected Green's functions on the right-hand side of Eq. (29). The
decompositions are easily performed with the help of the Legendre
transformation which is described by the relation between the generating
functional of proper vertices $\Gamma $ and the one for connected Green's
functions $W$ [16,17] 
\begin{equation}
\Gamma [A_\mu ^a,\overline{\psi },\psi ,\overline{C}^a,C^a]=W[J,\overline{%
\eta },\eta ,\overline{\xi },\xi ]-\int d^4x[J^{a\mu }A_\mu ^a+\overline{%
\eta }\psi +\overline{\psi }\eta +\overline{\xi }C+\overline{C}\xi ] 
\eqnum{A11}
\end{equation}
and the relations between the field functions and the external sources 
\begin{equation}
\psi (x)=\frac{\delta W}{\delta \overline{\eta }(x)},\overline{\psi }(x)=-%
\frac{\delta W}{\delta \eta (x)},A_\mu ^a(x)=\frac{\delta W}{\delta J^{a\mu
}(x)},C^a(x)=\frac{\delta W}{\delta \overline{\xi }^a(x)},\overline{C}^a(x)=-%
\frac{\delta W}{\delta \xi ^a(x)}  \eqnum{A12}
\end{equation}
\begin{equation}
\eta (x)=-\frac{\delta \Gamma }{\delta \overline{\psi }(x)},\overline{\eta }%
(x)=\frac{\delta \Gamma }{\delta \psi (x)},J_\mu ^a(x)=-\frac{\delta \Gamma 
}{\delta A^{a\mu }(x)},\xi ^a(x)=-\frac{\delta \Gamma }{\delta \overline{C}%
^a(x)},\overline{\xi }^a(x)=\frac{\delta \Gamma }{\delta C^a(x)}  \eqnum{A13}
\end{equation}
where the field functions in Eq. (A12) are all functionals of the external
sources in Eq. (A13) and, simultaneously, the sources in Eq. (A13) are all
functionals of the field functions in Eq. (A12).

Taking the derivative of the both sides of the first equality in Eq. (A12)
with respect to $\psi (y)$ and employing the first relation in Eq. (A13),
one may get 
\begin{equation}
\int d^4z\frac{\delta ^2\Gamma }{\delta \psi (y)\delta \overline{\psi }(z)}%
\frac{\delta ^2W}{\delta \eta (z)\delta \overline{\eta }(x)}=\int d^4z\frac{%
\delta ^2W}{\delta \overline{\eta }(x)\delta \eta (z)}\frac{\delta ^2\Gamma 
}{\delta \overline{\psi }(z)\delta \psi (y)}=-\delta ^4(x-y)  \eqnum{A14}
\end{equation}
where we only keep the term on the right-hand side of Eq. (A14) which is
nonvanishing when the sources are set to vanish. In order to find the
one-particle-irreducible decomposition for the quark-gluon three-point
Green's functions, one may differentiate Eq. (A14) with respect to the
source $J^{a\mu }(x_i)$ and then using Eq. (A14) once again. By this
procedure, it can be derived that 
\begin{equation}
\begin{tabular}{l}
$\frac{\delta ^3W}{\delta J^{a\mu }(x_i)\delta \overline{\eta }(x_j)\delta
\eta (y_k)}=\int d^4zd^4u_1d^4u_2\frac{\delta ^2W}{\delta J^{a\mu
}(x_i)\delta J^{b\nu }(u_1)}\frac{\delta ^2W}{\delta \overline{\eta }%
(x_j)\delta \eta (u_2)}$ \\ 
$\frac{\delta ^3\Gamma }{\delta A_\nu ^b(u_1)\delta \overline{\psi }%
(u_2)\psi (z)}\frac{\delta ^2W}{\delta \overline{\eta }(z)\delta \eta (y_k)}$%
\end{tabular}
\eqnum{A15}
\end{equation}
where the coordinates in Eq. (A14) have been appropriately changed. When all
the sources are set to be zero, noticing the definitions given in Eq. (A5)
where the $Z$ is replaced by $iW$ and in Eq. (A9) as well as 
\begin{equation}
\Delta _{\mu \nu }^{ab}(x_i-y_j)=\frac{\delta ^2W}{i^2\delta J^{a\mu
}(x_i)\delta J^{b\nu }(y_j)}\mid _{J=0}  \eqnum{A16}
\end{equation}
\begin{equation}
\Gamma ^{b\nu }(u_1\mid u_2,z)=i\frac{\delta ^3\Gamma }{\delta A_\nu
^b(u_1)\delta \overline{\psi }(u_2)\delta \psi (z)}\mid _{A=\overline{\psi }%
=\psi =0}  \eqnum{A17}
\end{equation}
the decomposition shown in Eqs. (30) and (31) straightforwardly follows from
Eq. (A15). Analogously, if we replace $\overline{\eta }(x_j)$ and $\eta
(y_k) $ by $\overline{\eta }^c(x_j)$ and $\eta ^c(y_k)$ in Eq. (A15) and
noticing 
\begin{equation}
\Gamma _c^{b\nu }(u_1\mid u_2,z)=i\frac{\delta ^3\Gamma }{\delta A_\nu
^b(u_1)\delta \overline{\psi }^c(u_2)\delta \psi ^c(z)}\mid _{A=\overline{%
\psi }=\psi =0}  \eqnum{A18}
\end{equation}
the decomposition shown in Eq. (32) and (33) will be derived. This
decomposition may also be derived from Eq. (A15) by the charge conjugation
transformation for the quark fields.

The one-particle-irreducible decomposition of the connected Green's function 
$G_c(x_{1,}x_2;y_1,y_2)$ can be derived by the same procedure as obtaining
Eq. (A15). On differentiating Eq. (A14) with respect to $\overline{\eta }%
^c(x_2)$ and $\eta ^c(y_2)$ and setting all the sources but the source $J$
to vanish, one may obtain 
\begin{equation}
\begin{tabular}{l}
$G_c(x_{1,}x_2;y_1,y_2)^J=\int
\prod\limits_{i=1}^2d^4u_id^4v_iS_F(x_1-u_1)^JS_F^c(x_2-u_2)^J$ \\ 
$\times \Gamma (u_1,u_2;v_1,v_2)^JS_F(v_1-y_1)^JS_F^c(v_2-y_2)^J$%
\end{tabular}
\eqnum{A19}
\end{equation}
where the four-point connected Green's function and the propagators given in
presence of the sources were defined before and the function $\Gamma
(u_1,u_2;v_1,v_2)^J$ is formally the same as that defined in Eqs. (38)and
(39). When the source $J$ is turned off, Eq. (A19) directly goes over to the
decomposition in Eq. (37)-(39) being defined in Eqs. (A17) and (A18) and in
the following: 
\begin{equation}
\Gamma _2(u_1,u_2;v_1,v_2)=i\frac{\delta ^4\Gamma }{\delta \overline{\psi }%
(u_1)\delta \overline{\psi }^c(u_2)\delta \psi (v_1)\delta \psi ^c(v_2)}\mid
_{\overline{\psi }=\psi =\overline{\psi }^c=\psi ^c=0}  \eqnum{A20}
\end{equation}
which is the quark-antiquark four-line proper vertex. It is emphasized here
that the decomposition of the function $G_c(x_{1,}x_2;y_1,y_2)$ in absence
of the source $J$ has the same form as that given in the presence of $J$.
This is because the Green's function is defined only by the differentials
with respect to the fermion fields as indicated in Eq. (A4).

The one-particle irreducible decomposition of the Green's function $G_{c\mu
}^a(x_i\mid x_{1,}x_2;y_1,y_2)$ may be derived by starting from the
expression given in Eq. (A15) with $j,$ $k=1$. By differentiating the both
sides of Eq. (A15) with respect to the sources $\overline{\eta }^c(x_2)$ and 
$\eta ^c(y_2)$ and then turning off all the external sources, one may obtain
the decomposition of the function $G_{c\mu }^a(x_i\mid x_{1,}x_2;y_1,y_2)$
as shown in Eqs. (40)-(48). Alternatively, the decomposition may also be
obtained by starting with the expression written in Eq. (A19). Substituting
Eq. (A19) into Eq. (A8), then completing the differentiation with respect to
the source $J^{a\mu }(x_i)$ and finally setting the source to vanish, one
may also derive the irreducible decomposition of the function $G_{c\mu
}^a(x_i\mid x_{1,}x_2;y_1,y_2)$. In doing this, it is necessary to perform
the differentiations of the fermion propagators with respect to the source $%
J^{a\mu }(x_i)$ as shown in Eqs. ( A9) and (A10) and use their
decompositions presented in Eqs. (30)-(33). In addition, we need to carry
out the differentiations of the gluon propagator and some vertices with
respect to the source $J^{a\mu }(x_i)$ as shown below. For the gluon
propagator defined in Eq. (A16), from its representation in presence of the
external source $J$, in the same way as deriving the decomposition
represented in Eqs. (A15), ( 30) and (32), one may obtain the one-particle
irreducible decomposition of the gluon three-point Green's function as
follows: 
\begin{equation}
\Lambda _{\mu \rho \sigma }^{acd}(x_i,z_1,z_2)=\frac \delta {\delta J^{a\mu
}(x_i))}\Delta _{\rho \sigma }^{cd}(z_1-z_2)^J\mid _{J=0}=\int d^4zD_{\mu
\nu }^{ab}(x_i-z)\Pi _{\rho \sigma }^{bcd,\nu }(z,z_1,z_2)  \eqnum{A21}
\end{equation}
where $D_{\mu \nu }^{ab}(x_i-z)\ $was defined in Eq. (34) and $\Pi _{\rho
\sigma }^{bcd,\nu }(z,z_1,z_2)$ was represented in Eq. (47) with the gluon
three-line proper vertex defined by 
\begin{equation}
\Gamma _{bcd}^{\nu \rho \sigma }(z,u_1,u_2)=i\frac{\delta ^3\Gamma }{\delta
A_\nu ^b(z)\delta A_\rho ^c(u_1)\delta A_\sigma ^d(u_2)}\mid _{A=0} 
\eqnum{A22}
\end{equation}
For other proper vertices related to fermions, we use the notation $\Gamma
_{\nu \lambda \cdot \cdot \cdot }^{bc\cdot \cdot \cdot }(z_1,z_2\cdot \cdot
\cdot \mid x_1,x_2\cdot \cdot \cdot )$ to represent them where color and
Lorentz indices and the coordinates on the left-hand side of the vertical
line belong to gluons and the coordinates on the right-hand side of the
vertical line are attributed to fermions, their derivative with respect to
the source $J^{a\mu }(x_i)$ can be represented as 
\begin{equation}
\frac \delta {i\delta J_\mu ^a(x_i)}\Gamma _{\lambda \tau \cdot \cdot \cdot
}^{cd\cdot \cdot \cdot }(z_1,z_2\cdot \cdot \cdot \mid x_1,x_2\cdot \cdot
\cdot )^J\mid _{J=0}=\int d^4zD_{ab}^{\mu \nu }(x_i-z)\Gamma _{\nu \lambda
\tau \cdot \cdot \cdot }^{bcd\cdot \cdot \cdot }(z,z_1,z_2\cdot \cdot \cdot
\mid x_1,x_2\cdot \cdot \cdot )  \eqnum{A23}
\end{equation}
where 
\begin{equation}
\Gamma _{\nu \lambda \tau \cdot \cdot \cdot }^{bcd\cdot \cdot \cdot
}(z,z_1,z_2\cdot \cdot \cdot \mid x_1,x_2\cdot \cdot \cdot )=\frac \delta {%
\delta A^{b\nu }(z)}\Gamma _{\lambda \tau \cdot \cdot \cdot }^{cd\cdot \cdot
\cdot }(z_1,z_2\cdot \cdot \cdot \mid x_1,x_2\cdot \cdot \cdot )^J\mid _{J=0}
\eqnum{A24}
\end{equation}
Especially, for the differential of the vertex $\Gamma (u_1,u_2;v_1,v_2)$ in
Eq. (38) with respect to the source $J^{a\mu }(x_i)$, we write 
\begin{equation}
\overline{\Gamma }_\mu ^a(x_i\mid u_1,u_2;v_1,v_2)=\frac \delta {i\delta
J^{a\mu }(x_i)}\Gamma (u_1,u_2;v_1,v_2)^J\mid _{J=0}  \eqnum{A25}
\end{equation}
Similarly, we define 
\begin{equation}
\overline{\Gamma }_{\mu \nu }^{ab}(x_1,x_2\mid u_1,u_2;v_1,v_2)=\frac{\delta
^2}{i\delta J^{a\mu }(x_1)i\delta J^{b\nu }(x_2)}\Gamma
(u_1,u_2;v_1,v_2)^J\mid _{J=0}  \eqnum{A26}
\end{equation}
the explicit expression given by the above differentiation can be obtained
by substituting Eqs. (38), (39) and (A20) into Eq. ( A25) and making use of
the formulas in Eqs. (A21) and (A23). The results are shown in Eqs.
(44)-(48). According to the procedure stated above, it is not difficult to
derive the expressions described in Eqs. (79)-(89). In the expressions, the
vertices are defined as in Eq. (A24). Some examples are listed below,

\begin{equation}
\Gamma _{\mu \nu }^{ab}(z_1,z_2\mid u_1,v_1)=i\frac{\delta ^4\Gamma }{\delta
A^{a\mu }(z_1)\delta A^{b\nu }(z_2)\delta \overline{\psi }(u_1)\delta \psi
(v_1)}\mid _{A=\overline{\psi }=\psi =0}  \eqnum{A27}
\end{equation}
\begin{equation}
\Gamma _{{\bf c}\mu \nu }^{ab}(z,z_1\mid u_1,v_1)=i\frac{\delta ^4\Gamma }{%
\delta A^{a\mu }(z_1)\delta A^{b\nu }(z_2)\delta \overline{\psi }%
^c(u_1)\delta \psi ^c(v_1)}\mid _{A=\overline{\psi }^c=\psi ^c=0} 
\eqnum{A28}
\end{equation}

\begin{equation}
\Gamma _\mu ^a(z\mid u_1,u_2;v_1,v_2)=i\frac{\delta ^4\Gamma }{\delta
A^{a\mu }(z)\delta \overline{\psi }(u_1)\delta \overline{\psi }^c(u_2)\delta
\psi (v_1)\delta \psi ^c(v_2)}\mid _{A=\overline{\psi }=\psi =\overline{\psi 
}^c=\psi ^c=0}  \eqnum{A29}
\end{equation}
and 
\begin{equation}
\Gamma _{\mu \nu }^{ab}(z_1,z_2\mid u_1,u_2;v_1,v_2)=i\frac{\delta ^4\Gamma 
}{\delta A^{a\mu }(z_1)\delta A^{b\nu }(z_2)\delta \overline{\psi }%
(u_1)\delta \overline{\psi }^c(u_2)\delta \psi (v_1)\delta \psi ^c(v_2)}\mid
_{A=\overline{\psi }=\psi =\overline{\psi }^c=\psi ^c=0}.  \eqnum{A30}
\end{equation}


\section{References}

\begin{references}
\bibitem{}  [1] E. E. Salpeter and H. A. Bethe, Phys. Rev. 84 (1951) 1232.

[2]M. Gell-Mann and F. E. Low, Phys. Rev. 84 (1951) 350.

[3]N. Nakanishi, Prog. Theor. Phys. Suppl. 42 (1959) 1 ; Prog. Theor. Phys.
Suppl. 95 (1988) 1,\ A great deal of references are cited therein.

[4]E. E. Salpeter, Phys. Rev. 87 (1952) 328.

[5]A. A. Logunov and A. N. Tavkhelidze, Nuovo Cimento 29 (1963) 380.

[6]Y. A. Alessandrini and R. L. Omnes, Phys. Rev. 139 (1965) 167.

[7]R. Blankenbecler and R. Sugar, Phys. Rev. 142 (1966) 1051.

[8]F. Gross, Phys. Rev. 186 (1969) 1448.

[9]I. T. Todorov, Phys. Rev. D3 (1971) 2351.

[10J. Bijtebier and J. Broekaert, J. Phys. G: Nucl. Part. Phys. 22 (1996)
559.

[11]J. Bijtebier, Nuovo Cimento A 100 (1988) 91; A102 (1989) 1235; A103
(1990) 317; 639; 669.

[12]W. Lucha, F. F. Sch\"oberl and D. Gromes, Phys. Rep. 200 (1991) 127,
many references concerning the $q\overline{q}$ bound state B-S equation can
be found therein.

[13]K. Erkelenz, Phys. Rep. 13 (1974) 191.

[14]J. C. Su, Commun. Theor. Phys. 38 (2002) 433.

[15]J. C. Su, J. Phys. G: Nucl. Part. Phys. 30, 1309 (2004)

[16]C. Itzykson and J. B. Zuber, Quantum Field Theory, McGraw-Hill, New
York, 1980.

[17]E. S. Abers and B. W. Lee, Phys. Rep. C9 (1973) 1.

[18]H. Lehmann, Nuovo Cimento 11 (1954) 342.

[19]R. Ricken, M. Koll, D. Merten, B. Ch. Metsch and H. R. Petry, Eur. Phys.
J. A9, 221 (2000).

[20]F. J. Dyson, Phys. Rev. 75 (1949) 1736.

[21]J. S. Schwinger, Proc. Nat. Acad. Sc. 37 (1951) 452, 455; Phys. Rev. 125
(1962) 397; 128 (1962) 2425.

[22]C. D. Roberts and A. G. Williams, Prog. Part. Nucl. Phys. Vol. 33, 477
(1994).
\end{references}
\end{document}